\documentclass[11pt, a4paper]{revtex4}

\usepackage{graphicx}
\usepackage{dcolumn}
\usepackage{latexsym}
\usepackage{amsmath, amsthm, amssymb}
\usepackage{epsfig}
\usepackage{bm}
\usepackage[dvipsnames]{xcolor}
\usepackage{caption,subcaption}
\usepackage{hyperref}
\usepackage{float}
\usepackage{aliascnt}
\usepackage{xcolor}
\usepackage{amsmath}
\newaliascnt{eqfloat}{equation}
\newfloat{eqfloat}{h}{eqflts}
\floatname{eqfloat}{Equation}

\newcommand*{\ORGeqfloat}{}
\let\ORGeqfloat\eqfloat
\def\eqfloat{%
  \let\ORIGINALcaption\caption
  \def\caption{%
    \addtocounter{equation}{-1}%
    \ORIGINALcaptionhttps://www.overleaf.com/project/6202ccc7b41cfdc6836afb8c
  }
  \ORGeqfloat
}
\usepackage[utf8]{inputenc}
\DeclareUnicodeCharacter{2212}{-}

\begin{document}

\title{Active Brownian particles can mimic the pattern of the substrate}
\author{Pawan Kumar Mishra}
\email{pawankumarmishra.rs.phy19@itbhu.ac.in}
 \altaffiliation{Department of Physics, Indian Institute of Technology (BHU), Varanasi, U.P. India - 221005}

\author{Ajeya Krishna}
\email{ajeya.krishnap.phy17@iitbhu.ac.in}
\affiliation{ 
Department of Physics, Indian Institute of Technology (BHU), Varanasi, U.P. India - 221005}
 \altaffiliation{SaaS Labs, Noida, Uttar Pradesh 201301}
 
\author{Shradha Mishra}%
 \email{smishra.phy@itbhu.ac.in}
\affiliation{ 
Department of Physics, Indian Institute of Technology (BHU), Varanasi, U.P. India - 221005
}%

\begin{abstract}
\section{Abstract}
Active Brownian particles (ABPs) are termed out to be a successful way of modeling the moving microorganism on the substrate. In recent studies, it is shown that such organisms can sense the characteristics of the substrate. Motivated by such work, we studied the dynamics and the steady state of ABP moving on a substrate with space-dependent activity. On the substrate, some regions are marked as high in activity, and other regions are such that particles behave as passive Brownian particles. The system is studied in two dimensions with step, sigmoid, Gaussian and cone shape distribution of activity profile on the substrate.  The whole interface of the activity profile is symmetrically divided into two regions. This lead to the flow of particles from the active region to the passive region. The final steady state of particle density profile, polarisation and flux very much follows the structure of the inhomogeneous activity and the density in high activity region is lower, maximum at the interface and nearly constant with mean density in the passive region. Further, the steady state density profile for various shapes and designs on two-dimensional substrates. Hence the collection of ABPs on an inhomogeneous substrate can mimic the inhomogeneity of the substrate.

\end{abstract}

\maketitle

\section{Introduction}
Active systems 
\cite{model_h,model_b} have been a frontier area of research
in the last two decades because of their unusual properties as compared to the systems in thermal equilibrium.  
The rapidly growing discipline of active matter\cite{active_1,active_2,active_3,active_4,active_5,active_6,active_7} 
seeks to understand and regulate the material characteristics of assemblages of 
interacting energy-consuming \cite{energy} components on a microscopic level. 
Examples of motile active matter may be found everywhere in nature, 
from flocks of birds \cite{birds} to family of fish \cite{fish} to bacterium colonies like { \textit{Escherichia coli}} \cite{bacteria}. 
Many laboratory investigations of artificial active fluids of suspended lifeless microswimmers have been prompted by the abundance of natural occurrences observed. 
Simple colloidal particles driven by self-phoresis \cite{p_1,p_2,p_3,p_4,p_5,p_6,p_7,p_8} are frequently used in these ``active-particle systems" \cite{allanpnas2017}. The individual constituents of these systems transduce their internal energy into motion, i.e., they exhibit self-propulsion characteristics, and therefore, they are also called self-propelled
particles (SPPs).
On the level of a single or a few active particles \cite{p_3},\cite{ap_1,ap_2,ap_3,ap_4,ap_5,ap_6}, several intriguing characteristics have been identified, opening up a wide range of possible applications. 
Microswimmers also have a diverse range of collective dynamics, ranging from mesoscopic turbulence \cite{turbulance} to macroscopic motility-induced phase separation (MIPS) \cite{mips_1,mips_2,mips_3,mips_4,mips_5,mips_6,mips_7,mips_8,mips_9,mips_10}. In addition to the extensive study of these systems in clean environments \cite{s1,s2, abp_1, Vicsek1995, chate2007comment,chatepre2008,Toner1995,tonertupre1998}, recently people have started to look for their bulk properties in
heterogeneous medium \cite{quint2015topologically,morin2017distortion, chepizhko2013optimal,rakeshPRE,Toner2018pre, Toner2018prl,Peruani2018prl,sandor2017pre, sandor2017jcp}.\\
Inhomogeneity can exists at all scales of active matter.
The active matter community has only lately concentrated its attention on a detailed examination of typical patterns in local density and polarisation in
the inhomogenoues conditions \cite{ref, preinhomogeneous2017}. In recent study \cite{hu2015physical} it is shown that the microswimmers can sense the geometry of the substrate and hence can be used as biosensor.

Also, the recent experiment \cite{Choudhury_2017} and simulation  \cite{semwal2021dynamics}, consider the motion of the active {Brownian} particles
moving on the top of two-dimensional periodic surface. It is found  that the ABPs  experiences competition between hindered and enhanced diffusion due the pattern of the surface.
In these studies the inhomogeneity is introduced as different surface morphology. 
Here our aim is to exmine whether the inhomogeneity in activity profile will affect the collective properties of ABPs. 
In this work  we  focus on studying the behaviour of collection of ABP's moving on inhomogeneous substarte of  various distributions of activity profile. We studied the  various distributions  using the {\em coarse-grained} as well as {\em microscopic simulations}  of  active Brownian particles (ABPs) \cite{model_b}.\\
The response of real microswimmers to the local profile of activity can be experimentally designed by introducing different thermal gradient, light induced activity profile or field induced \cite{abdoli2021stochastic, stenhammar2016light, vuijk2020lorentz, vuijk2022active}. Also it can be naturally present in the systems where bacteria is moving in the background polymer network as discussed in review work of \cite{cates2015motility}.  The above works show the  strong dependence of steady state density of microswimmers in response to the activity profile. In addition to that in our present work we also show that the current and density of particles very much depends on the gradient of the activity profile.


In this study, we  consider a collection of active Brownian particles (ABPs) \cite{mips_5,abp_1,abp_2,abp_3,abp_4} moving on a two-dimensional inhomogeneous substrate. The substrate is designed such that the particle experiences high activity in some regions and  no or small  activity at other places. We have considered varied distributions of activity for the substrate. They are of type; Step function, Sigmoid, Gaussian, Cone and random shape distributions. Our main result is that  the particle density can mimic the underneath substrate structure.\\
The rest of the article is divided in the following manner. In the next section  \ref{model}, we discuss the model in detail and finally discuss the results in section \ref{results}. In section \ref{random} we discuss the steady state density profile for various {miscellaneous shapes}.\\


\section{Model}
\label{model}
We consider a collection of active Brownian particles (ABPs) moving on a two-dimensional inhomogeneous substrate. The substrate is designed such that
there are regions, where particle experiences high activity and regions where  it behave as passive particle. Hence particles move on a substrate with inhomogeneous
activity profile.  The  dynamics of ABP's is given by the following update equations for the particles position on  the two-dimensional substrate;
\begin{equation}
    \partial_tx = v(x,y)cos\theta + \sqrt{2D\xi_x}
    \label{eq : 1}
\end{equation}
\begin{equation}
    \partial_ty = v(x,y)sin\theta + \sqrt{2D\xi_y}
    \label{eq : 2}
\end{equation}
\begin{equation}
    \partial_t\theta = \sqrt{2D_r\xi_\theta}
    \label{eq : 3}
\end{equation}
The first term on the  right hand side of the both Eq.s \ref{eq : 1} and  \ref{eq : 2} incorporates the inhomogeneous activity with space dependence of self-propulsion speed $v(x, y)$. The $\xi_{i}$'s are due to  the  thermal noise (translational and rotational).
The transitional and rotational diffusion coefficients $D$
and $D_r$, respectively, measure intensities of independent,
unit variance,  Gaussian white noises
$\xi_{x,y,\theta}(t)$.

In the following section, we utilize the framework of
ref. \cite{ref} to derive approximate differential equations for
the  density ${\rho(r, t)}$ and local polarisation ${\bf p}(r, t)$  
at position $r$ and time $t$. The collection of ABPs governed by the Eq. \ref{eq : 1} to \ref{eq : 3}  is called as {\em microscopic} ($MIC$) model. Later we also write the coarse-grained coupled equations for local density $\rho(r, t)$ and polarisation ${\bf p}(r, t)$.

\subsection{Moment Equations}
The dynamic probability density function $f(r,\hat{n},t)$ for finding the particle  at time $t$ at position $r$ with the orientation $\hat{n} = (cos\theta,sin\theta)$, corresponding to the system of stochastic differential Eq. \ref{eq : 1},\ref{eq : 2} and \ref{eq : 3}, obeys the Fokker 
Planck Eq.\cite{FPE_1, FPE_2, FPE_3}

\begin{equation}
    \frac{\partial f}{\partial t} = D \nabla^2 f + D_r \partial_\theta^2f - \nabla.[fv(r)\hat{n}]
    \label{eq : 4}
\end{equation}

here, $\frac{\partial}{\partial t}$,  $\nabla$ and $\partial_{\theta}^2$ represents differential operators with  respect to time $t$ and space ${\bf r}$ and angle $\theta$ respectively. 
The exact moment expansion of $f$ in  terms  of $\hat{n}$ \cite{FPE_1,FPE_3,hat_n} truncated after the second term leads
\begin{equation}
    f(r,\hat{n},t) = \frac{1}{2\pi}[\nabla(r,t) + 2 {\bf p}(r,t) \cdot \hat{n}]
    \label{eq : 5}
\end{equation}
where,
\begin{equation}
    \rho(r,t) = \int d\hat{n}f(r,\hat{n},t),
    \label{eq : 6}
\end{equation}
\begin{equation}
    {\bf p}(r,t)  = \int d\hat{n}\hat{n}f(r,\hat{n},t)
    \label{eq : 7}
\end{equation}

denote  density and polarization, respectively. Multiplying Eq. \ref{eq : 4} by $1$ or $\hat{n}$, 
integrating over orientational degrees of freedom, and using the definitions \ref{eq : 6}, \ref{eq : 7}, 
we obtain the moment Eq.\cite{ABP,FPE_3} 

\begin{equation}
	\partial_t \rho(r,t) = -\nabla \cdot {\bf J}(r,t),
    \label{eq : 8}
\end{equation}
\begin{equation}
	\partial_t {\bf p}(r,t) = -D_r {\bf p}(r,t) - \nabla \cdot {\bf M}(r,t)
    \label{eq : 9}
\end{equation}
here, we introduced the (orientation averaged) flux \cite{main}
\begin{equation}
	{\bf J}(r,t) \equiv -D \nabla \rho(r,t) + v(r){\bf p}(r,t)
    \label{eq : 10}
\end{equation}
and the matrix flux
\begin{equation}
	{\bf M}(r,t) \equiv -D \nabla {\bf p}(r,t) + \frac{v(r)}{2}\rho(r,t) \bf{I}
    \label{eq : 11}
\end{equation}

with the unit matrix ${\bf I}$. The set of equations given by Eq.\ref{eq : 8}-\ref{eq : 11} is named as {\em coarse-grained} (CG) model.
\\

\bf{Numerical Details} : \normalfont{We have performed numerical integration of CG model using Euler's numerical integration method to solve eqs. \ref{eq : 8}, \ref{eq : 9}, \ref{eq : 10}, \ref{eq : 11}. We randomly initialized density $\rho$ and polarization ${\bf p}$ and calculated the orientation averaged flux ${\bf J}$ as given in eq. \ref{eq : 10} and matrix flux ${\bf M}$ as given in eq. \ref{eq : 11}. Then the density $\rho$ is updated by inserting ${\bf J}$ into eq. \ref{eq : 8} and similarly the polarization ${\bf p}$ is updated by inserting ${\bf M}$ into eq. \ref{eq : 9}. {We define the smallest length scale as $l= \frac{v_o}{D_r}$ and time scale $\tau = {D_r}^{-1}$. All the lengths and times are measured in multiple of them. The step size we considered are $dx = dy = 1.0 l$ and $dt = 0.01 \tau$. All the results are also checked for smaller step size $dx = dy = 0.5$ and we found the same results. Hence there is no ambiguity with respect to step size of integration and discontinuity of the inhomogeneous profile.} We started with  mean value of density $\rho_0 = 0.5$ and it remains constant throughout the simulation. 
The system is studied in two-dimensions. We first give the  details of parameter for CG model. We considered the length of the system $L=512 l$. The total simulation time is $2 \times 10^5 \tau$.  The periodic boundary conditions (PBC) is used in both directions. The inhomogeneous activity on the substrate is introduced through following four symmetric shapes of activity on the substrate with the maximum intensity, $v_o$ of the distribution: \\
(i) {\em Step Function Shape}:  We first introduced the step function  distribution of activity on the substrate such that
\begin{equation}
     v(x,y) = \begin{cases} 
      v_o & \sqrt{((x-\mu)^2+(y-\mu)^2)}  \leq r_o \\
      $0$ & \sqrt{((x-\mu)^2+(y-\mu)^2)} > r_o \\
    \end{cases}
   \label{step}
\end{equation}

We have a constant value for activity inside the circular region of radius $r_o$ = $\frac{1}{4} L$ and discontinuous 
at the boundary between the active and inactive regions in the step distribution.  \\

(ii) {\em Sigmoid Shape}:
The two-dimensional sigmoid function is defined as:
\begin{equation}
v(x, y) = v_0\times(1 - \frac{1}{1 + e^{-k \cdot \sqrt{(x - \mu)^2 + (y - \mu)^2}}})
   \label{sigmoid}
\end{equation}
where $\mu$ is the center parameter and \( k \) is the steepness parameter that controls the shape of the sigmoid function. We fix $k=50$ throughout the simulation.
This is same as the step distribution. The only difference is  rapid, continuous fall in activity at the boundary. \\

(iii) {\em Gaussian Shape}: Gaussian distribution  with  parameters $\mu, \sigma, v_0$ \ref{eq:lambda} \cite{genz2009computation}.
\begin{equation}
	v(x,y) = v_o \times \frac{1}{2\pi\sigma^2} \exp(-\frac{1}{2}[(\frac{x-\mu}{\sigma})^2 + (\frac{y-\mu}{\sigma})^2])
    \label{eq:lambda}
\end{equation}

$\mu$ is the center of the system.  $\sigma = \frac{1}{4} L$ represents variance in both  directions. 
$x$ and $y$ in eq. \ref{eq:lambda}, represents $x$ and $y$ coordinates in the system. \\

(iv) {\em Cone Shape}:  We introduce the cone shape distribution of activity on the substrate such that
\begin{equation}
     v(x,y) = v_0 \times (1 - \sqrt{\left({x -\mu}\right)^2 + \left({y - \mu}\right)^2})
   \label{uniform1}
\end{equation}

where $\mu = L/2$ is at the  center of the box.
In eq. \ref{uniform1}, the activity falls linearly away from the center of substrate and finally becomes zero. 
 \\
Further we also perform  {\em microscopic} simulation of Eq. \ref{eq : 1} to \ref{eq : 3} for the  {\em step} and {\em Gaussian} profile of activity on a two dimensional substrate of size $L=128 l$ with the number of active Brownian particles $N=8692$ and their radii $R=0.6 l$ . The dynamics of the ABPs are governed by a discrete time step, dt, set to $0.01 \tau$ units. At each time step, the particles undergo self-propulsion with a speed denoted by $v_o$. The total simulation time  set to $5 \times 10^4 \tau$ units. The repulsive force acting between two particles is modeled using a soft-core spring potential 
\begin{equation}
    F_{\text{rep}}(r_{ij}) = -K \cdot (2 R - r_{ij}) \cdot \hat{r}_{ij}
   \label{F_rep}
\end{equation}
where $r_{ij}$ is the distance between particles $i$ and $j$, $K$ is a positive constant that determines the strength of the repulsive force and it is taken as unit in the simulation and $\hat{r}_{ij}$ is the vector pointing from particle $i$ to $j$.
The force is zero when the distance between particles exceeds the sum of their radii, and it increases as particles get closer. The last term in Eq.\ref{eq : 1} and \ref{eq : 2} is the {translational} noise which is present due to thermal fluctuations. The angular noise in eq.\ref{eq : 3} affects the rotational diffusion of particles and is responsible for the random changes in their orientations. The strengths of translations and rotational diffusion constants are fixed to $1.0$ for both.

The system is studied for $v_o=1.0$ in both CG grained and MIC simulations and rest of other parameters are given in each activity profiles. { The dimensionless maximum {\em Peclet} number $Pe = \frac{l}{R} =  1.33$.}


\section{Results}
\label{results}
In this section we discuss our results in detail. We divide our study in two parts. In first part we show the results of four different shapes of activity profile given in previous section using the CG simulations of Eq.\ref{eq : 8} to \ref{eq : 11}. We also discuss the time evolution of system towards the steady state for one shape of activity profile: {\em i.e.} Gaussian shape function. In the second part we show the time evolution of the system studied using the MIC model introduced in Eq.\ref{eq : 1}-\ref{eq : 3}. 

\subsection{Coarse-grained model (CG)}
We analyze the density ${\rho}$, polarization ${\bf p}$, and flux ${\bf J}$ (as given in Eq.\ref{eq : 8}, \ref{eq : 9} and \ref{eq : 10} respectively) of the particles observed for various shapes of speed distribution (activity profile) including step, sigmoid, Gaussian and cone shape in Fig.\ref{2dprofile}.

For the time at $t=200000$ and for system size $L=512$, we find that the density is least in the highly active region, reaches its maximum at the interface between the high and low active regions.  We can see in the Fig.\ref{2dprofile}, distribution of density depends on the local gradient of activity profile. Since the shape of step distribution of speed has very high gradient at the interface of active and passive region but zero gradient in the rest of the area,  we find that there is a sharp change from low density inside the active region to high density at the interface. When we introduce other activity profiles in the model we observe that there is a region of intermediate density between high and low density. This intermediate density is equal to the average density taken in the system. As we go from sharp to gradual (smooth) decay of gradient of  activity profiles such as sigmoid to Gaussian and finally to cone shape, the size of the intermediate region with high density  is increasing.
 The differences in the density profiles for different shapes highlight the influence of the activity profile on the system's behavior. Three dimensional representation of activity profile and corresponding density distribution is shown in Fig.\ref{3dpicture}. We observe depletion of the particle density from the active region and the area of the depleted region  is approximately equals the area of region where velocity is non-zero.\\
The corresponding polarization ${\bf p}$ in Eq.\ref{eq : 9} and current/flux ${\bf J}$ in Eq.\ref{eq : 10} in the steady state is shown in third and fourth row in Fig. \ref{2dprofile} indicate the magnitude and direction represented by color map and stream of arrows respectively . These two observables ${\bf p}$ and ${\bf J}$ are essential in understanding the overall dynamics in the system and how particles move and interact within the active region. The polarization is symmetric and points away from the centre of the substrate. Magnitude of polarization is maximum at the interface but the maximum value decreases for smooth activity profiles as shown in Fig. \ref{2dprofile} (Polarisation).

\begin{figure}[H]
  \includegraphics[width=16cm,height=16cm]{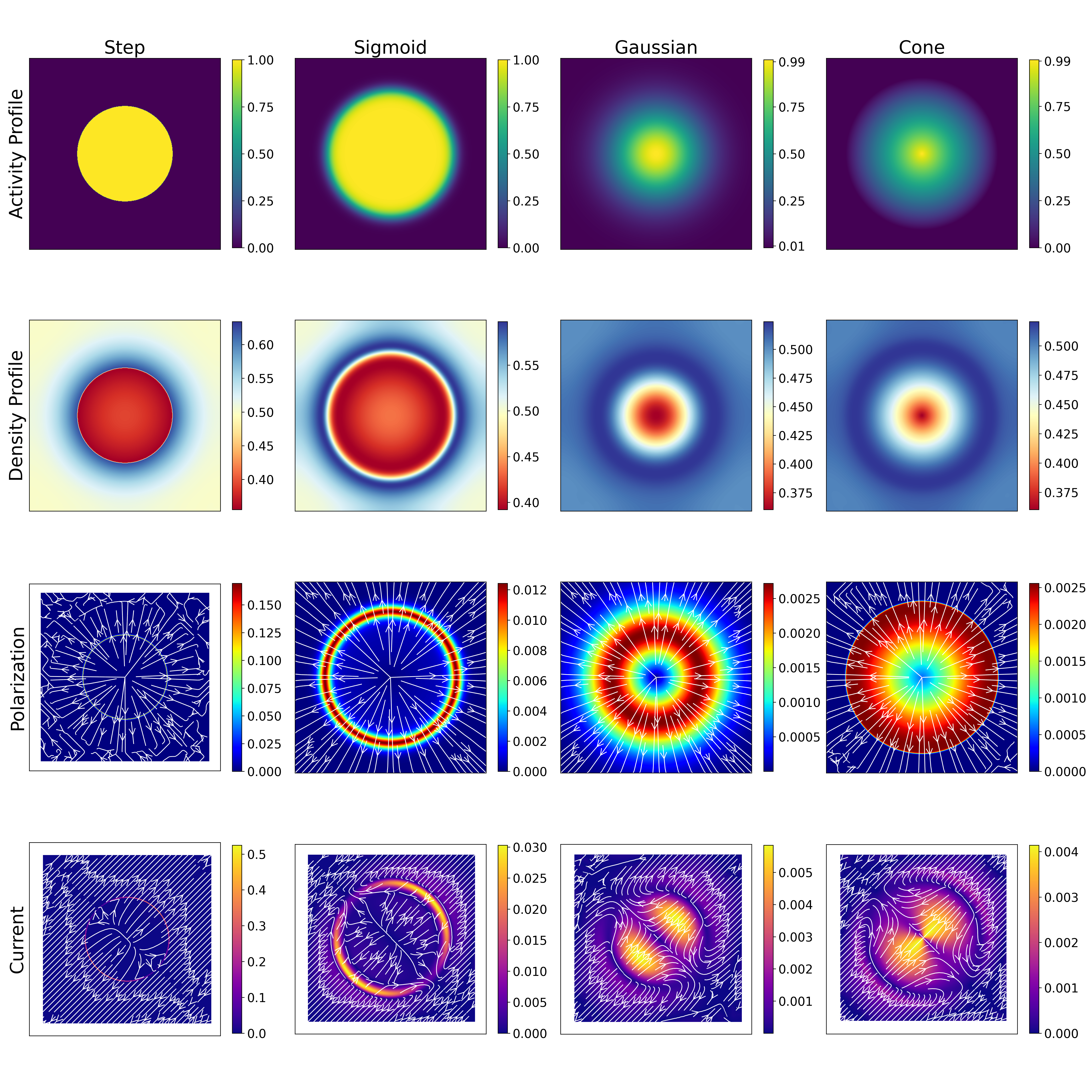}  
	\caption{The figure shows the activity profiles in the first row for different shapes of distributions of activity of Active Brownian Particles (ABPs). Each subplot in the columns represents a different shape of distribution: Step function shape, Sigmoid function shape, Gaussian function shape, and Cone shape from left to right respectively. The activity profiles are represented by color intensity, with warmer colors indicating higher activity levels and cooler colors indicating lower activity levels.
In the second, third, and fourth rows, the density $\rho({\bf r}, t)$, polarization ${\bf p}({\bf r}, t$), and flux/current ${\bf J}({\bf r}, t$) of the ABPs are displayed from top to bottom, respectively. The color intensity in the second row represents the density levels, with red indicating higher densities and blue indicating lower densities. In the third and fourth rows, the color intensity represents the magnitude of polarization and flux, with warmer colors indicating higher magnitudes and cooler colors indicating lower magnitudes.
The stream of arrows in the third and fourth columns illustrate the direction of polarization and flux, respectively. The rest of the system parameters are given in the main text.
}
\label{2dprofile}
\end{figure}
\

The current ${\bf J}$ is plotted in Fig.\ref{2dprofile} (Current).
We observe particles flow from high density (maximum current) region at interface to the low density regions both inward (high activity region) and outward (low activity region) but due to high activity they leave the active region in opposite directions.
The value of maximum current at the interface decreases for smooth activity profiles as represented by color bars. {The presence of angular dependence of current is solely due to the PBC in the system.} 

\begin{figure}[H]
  \includegraphics[width=16cm,height=8cm]{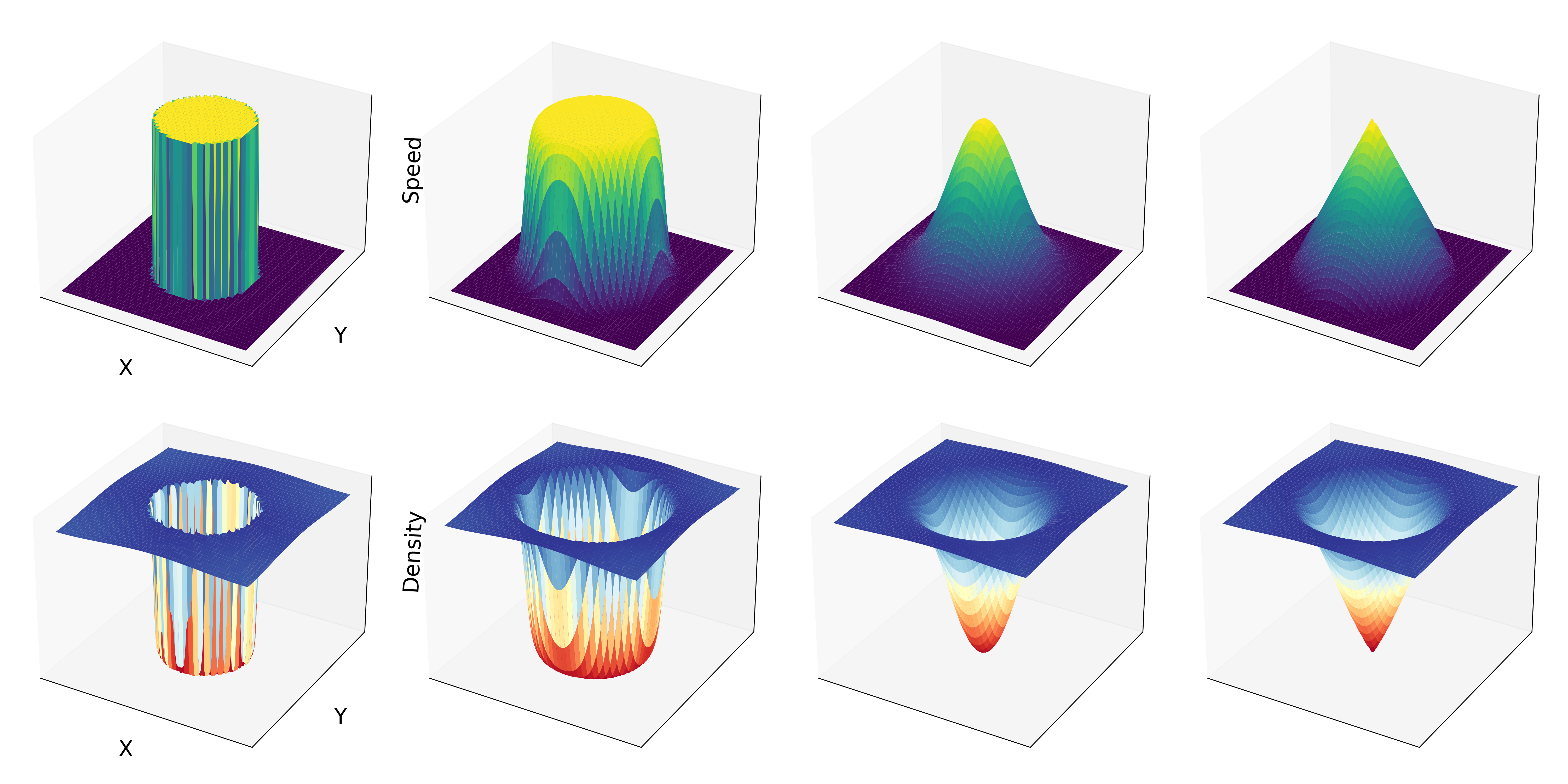}  
	\caption{This figure shows the three dimensional representation of step, sigmoid, Gaussian and cone shapes of activity profile in the first row. Activity profile is represented by speed of the particles on the {\em z}-axis on two dimensional {\em xy} plane The density corresponding to different shapes of activity profiles is shown on the {\em z}-axis in second row. These 3d snapshots are shown for system size, L=128 at time, t=10000. Note: that the density profile in the steady state exactly opposite to that of activity profile. }
\label{3dpicture}
\end{figure}
We show time evolution of density, polarisation and current for Gaussian shape of activity profile in Fig.\ref{timesnap}. We plot the color snapshots of particle density at different time steps from left to right $t_1$=1, $t_2$=1000, $t_3$=40000 and $t_4$=200000.   {Density pattern evolves with time} and we observed  the  depletion of particle density inside the active region. 
In the middle i.e., the area where the activity of the particles is maximum, we observe the lowest density of the particles and going from center to the boundary of the active and passive regions, we observe the density of the particles gradually increase. At the initial times, we observe the depletion of the particles getting bigger, but the depletion reaches a saturation state with time. \\
In the second row of the Fig. \ref{timesnap}, we observe the polarization snapshots of ${\bf p}$  with its magnitude represented by color map and direction with stream arrows. 
We observe at early times the particles are at random motion and orientation. Along with the time, we observe the polarization of particles increases in the active region. Except at the center of the system, we observe lower polarization. At later time there is not much change in the profile  with time. At early stages, there is no ordering of the polarization i.e., random direction of the polarization. With time we observe the formation of high polarization regions near the interface inside the high activity  region. At the center of the box, where the activity is the highest, we observe low polarization.\\
\begin{figure}[H]

  \includegraphics[width=16cm,height=10cm]{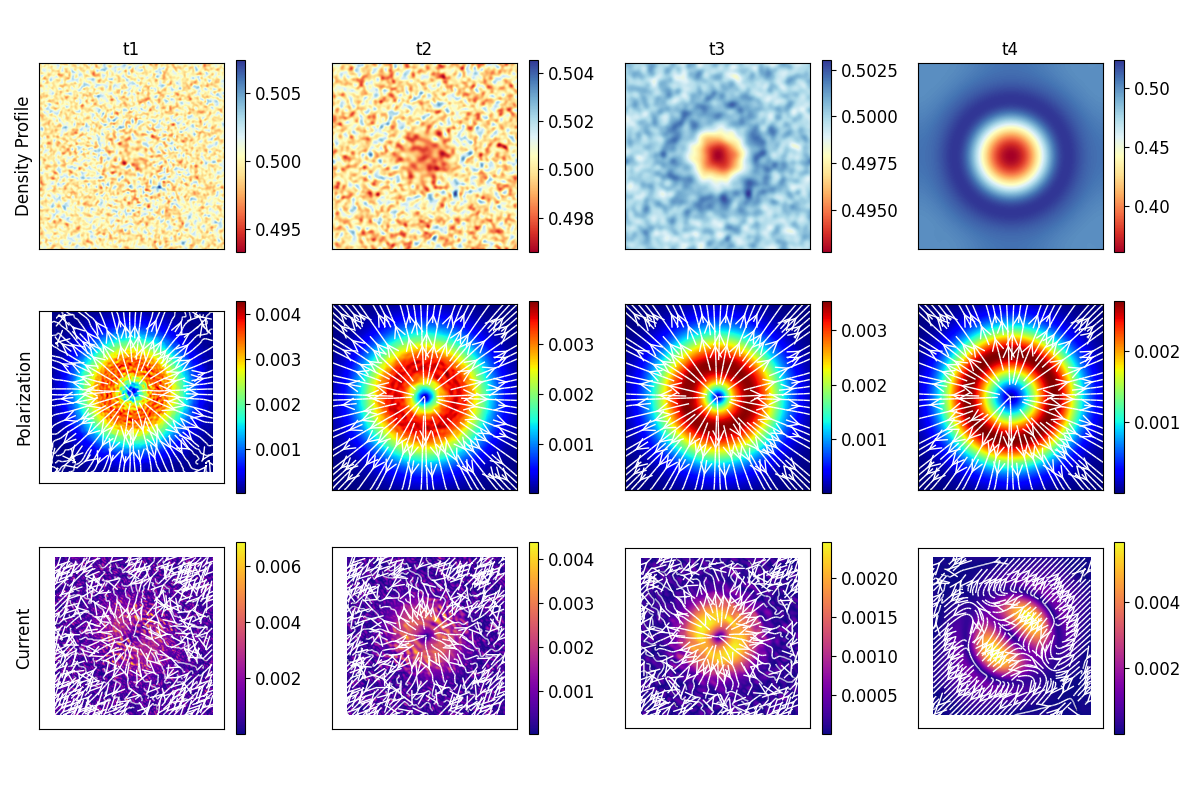}
  
	\caption{Time evolution snapshots of the model with Gaussian shape distribution of speed at time steps $t_1$=1, $t_2$=1000, $t_3$=40000 and $t_4$=200000 is shown from left to right. Density, polarisation and flux is shown in first, second and third row of the figure, respectively at the given times. The color bars represent the value of the density, polarisation and flux of the particles from low at bottom to high at the top. The stream of arrows represent the direction of polarisation and flux in the model.}
\label{timesnap}
\end{figure}

From the current plot in third row of Fig.\ref{timesnap}, we can see the current snapshots in the model. The color map represents the magnitude of the current and direction shown by stream arrows. In this model, at early times, we observe the particles are at random motion and orientation. With time, we see high current in the activity region.
 At later times as the system evolves, i.e., we don't observe major change in the density, polarisation and current profiles  with time. 
 
\begin{figure}[H]

  \includegraphics[width=16cm,height=8cm]{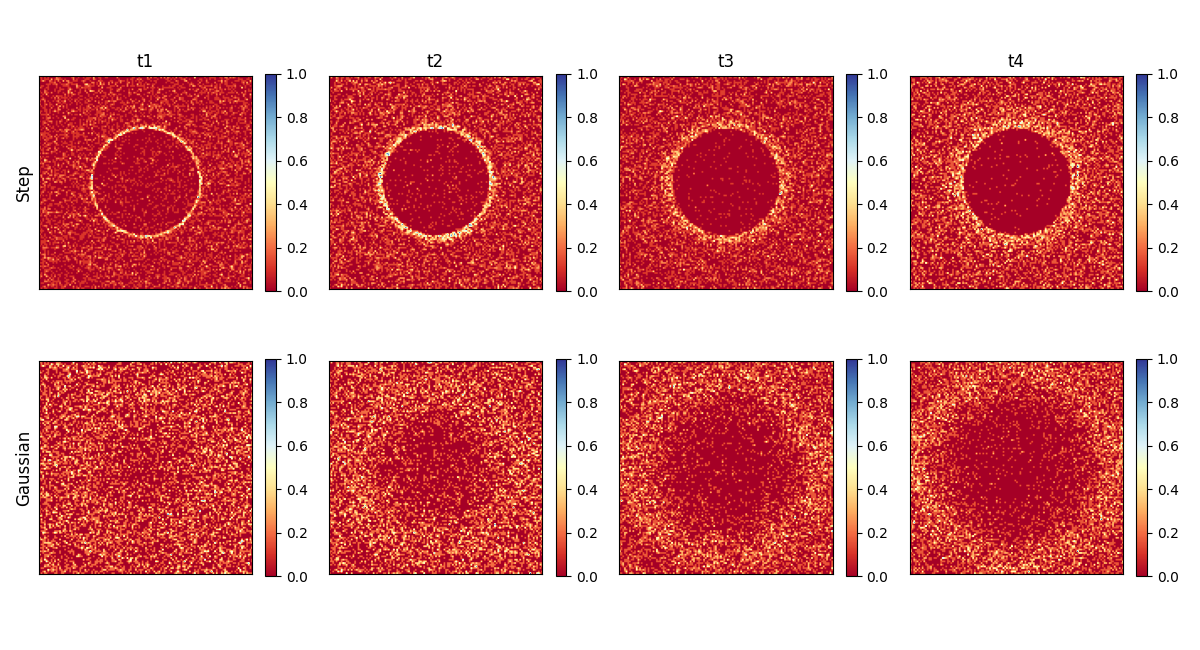}
  
	\caption{The figure illustrates the time-evolution of particle densities on substrates for the microscopic model $(MIC)$ for two different types of activity profiles i.e. step and Gaussian shape, represented in the top and bottom rows, respectively. Each column corresponds to a specific time point ($t_1=15000$, $t_2=30000$, $t_3=40000$, and $t_4=50000$). The particle density is visualized using color maps, with colors ranging from red to blue, indicating low to high density, respectively. }
\label{mictimesnap}
\end{figure}

\subsection{Microscopic model (MIC)}
Now we discuss the results of microscopic model introduced by Eq.\ref{eq : 1}-\ref{eq : 3}. We focus on the time evolution of local density of particles $\rho({\bf r}, t)$, where $\rho({\bf r}, t)$ is obtained by counting number of particles in unit square box, such that the whole system is divided in $L^2$ boxes of unit size. Here we only focus on the density profile of particles in the system for two activity profiles (i) step function and (ii) Gaussian. Fig. \ref{mictimesnap} shows the time evolution of local density at times $t_1=15000$, $t_2=30000$, $t_3=40000$, and $t_4=50000$. The system is started with random position and orientation of particles in whole substrate.  With time density $\rho({\bf r}, t)$ evolves such that inside the activity region less number of particles hence the lower density and outside or in the passive region almost mean density. At the interface we observe highest density of particles. The color bar in Fig. \ref{mictimesnap} shows the magnitude of local density and {\em note} that at the interface region density  is high. For step distribution the width of the high density interface is much sharper than that of the Gaussian case as shown in Fig.\ref{mictimesnap}.\\
{We demonstrate the remarkable capability of the Microscopic (MIC) model in capturing essential features of the system's behavior when compared to the Coarse-Grained (CG) model. In Fig. \ref{radialplot}, we present radial density profiles as a functions of radial distance from the center, with a specific focus on the activity profile of the step function.
The radial density profile $\rho(r)$ is calculated by  averaging of local density over  all  directions. 
The main plot illustrates the behavior of the density profile in the CG model, with a system size of $L=512l$ and its center located at coordinates $(256, 256)$. In this configuration, the interface of the activity profile is positioned at a radius $r=32 $.
Furthermore, the inset plot in Fig. \ref{radialplot} highlights the corresponding density profile in the MIC model, configured with a system size of $L=128 l$, and a center at coordinates $(64 l, 64 l)$. In this scenario, the interface of the activity profile also lies at radius $r=32$. This representation reveals the striking similarity in the radial variation of density between the CG and MIC models, particularly with respect to the step function activity profile. We observe for both cases that density profile shows a jump at interface and gradually  decreases to global mean density of the system. For MIC model packing fraction is $0.6$ and for CG model mean density is $0.5$.}

\begin{figure}[H]
   
\begin{center}
   \includegraphics[width=8cm,height=7cm]{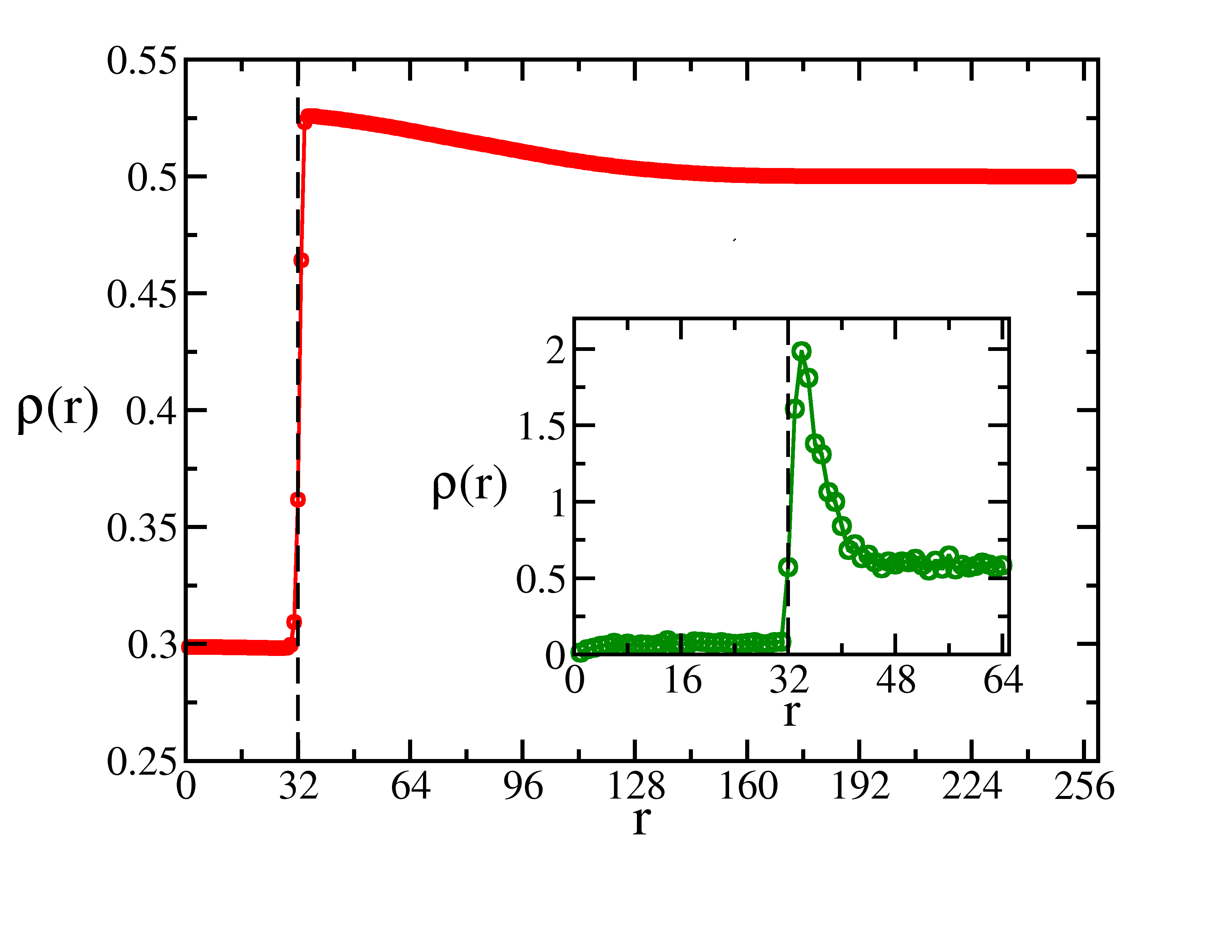}
  
	\caption{{Radial density as a function of distance from the center for the activity profile of step function, illustrating the behaviour of density profile in CG model (main plot) and the MIC model (inset plot).}   }
\label{radialplot}
 \end{center}
\end{figure}

\section{Miscellaneous shapes}
\label{random}
In previous two cases we have learnt that the particles deplete  from  the active region. 
The particles density can mimic the distributions underneath them. We ask the question; whether the above mechanism be recreated for some {miscellaneous} shapes without any symmetry. This is the motivation to consider {miscellaneous} shapes. The aim is to check whether the particles follow the pattern of the substrate as we had observed with various activity profiles such as step, sigmoid, Gaussian and cone shape distribution.{All the results of {miscellaneous} shapes are performed with step size $dx = dy = 1.0$, since we have checked the results for $dx = dy = 0.5$ for step profile, we believe that results of {miscellaneous} shapes will also remain unchanged for small step size.}
\begin{figure}[H]
\includegraphics[width=16cm,height=10cm]{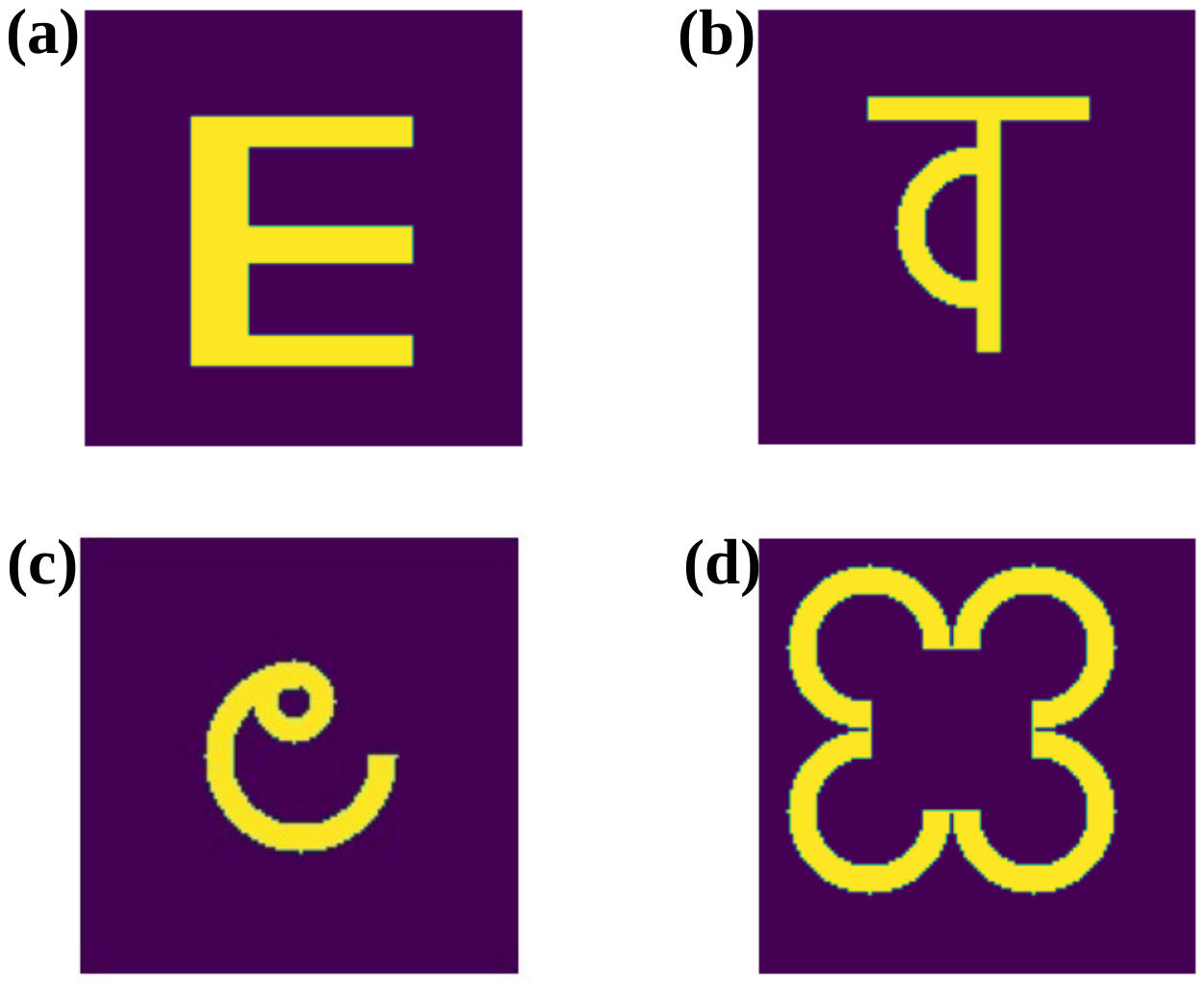}   

\caption{This figure shows the various distributions we have considered. a) E shaped potential \\
b) VA shaped potential c) LA shaped potential d) Flower shaped potential.}

\label{fig:5}
\end{figure}
 The various other distributions we looked are:

\begin{enumerate}
    \item English Alphabets
    \item Hindi Alphabets
    \item Telugu Alphabets
    \item Flower Shape
\end{enumerate}
In the Fig\ref{fig:5}, we observe four different shaped potentials. We actually had considered various other type potentials in these categories. In English Alphabets, we have considered Alphabets $E$, ($A$, $M$, $J$, $P$ (data not shown)) Fig.\ref{fig:5}(a). In Hindi Alphabets \cite{devnagri}, we have considered Alphabets $VA$, ($PA$, $NA$ (data not shown) Fig. \ref{fig:5}(b). In Telugu Alphabets \cite{telugu}, we have considered $LA$ Fig.\ref{fig:5}(c) and flower shape Fig.\ref{fig:5}(d). In these figures, the yellow color part is the active region and purple color part is the passive region. In the following section, we are attaching the steady state snapshots at time $t=100000$ of particles density distribution for different  distributions.

\begin{figure}[H]
    \centering 

\includegraphics[width=16cm,height=10cm]{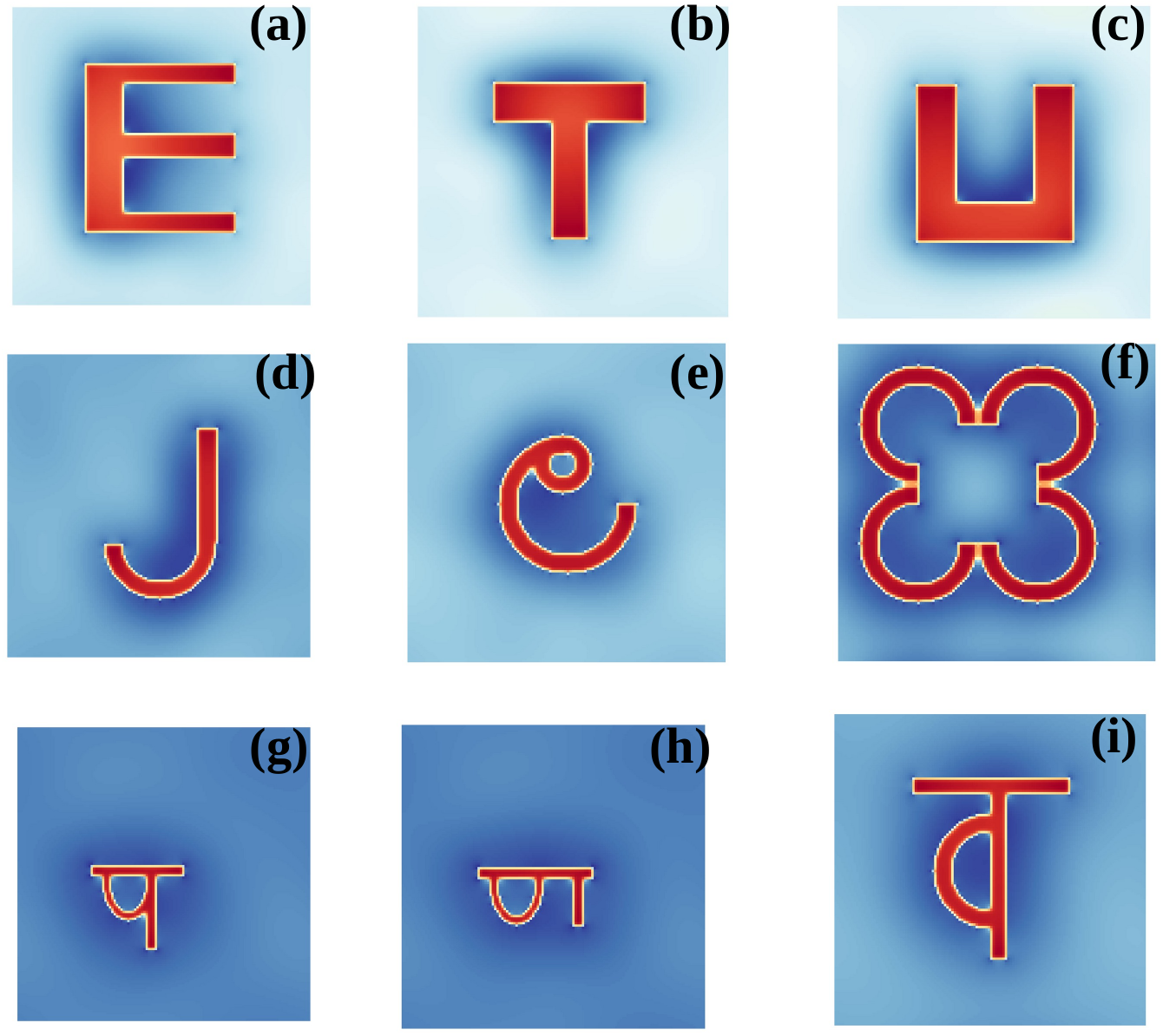}  

\caption{Steady state snapshot of the particles density distribution for different potentials. a) English Alphabet 'E' b) English Alphabet 'T' c) English Alphabet 'U' d) English Alphabet 'J' e) Telugu Alphabet\cite{telugu} 'la' f) random shape 'flower' g) Hindi Alphabet 'pa' h) Hindi Alphabet 'na' i) Hindi Alphabet 'va'. The red region represents low density and blue region represents high density of particles. All the snapshots are taken at an activity of $2.0$ and at the time t=50000. The box size is $128$ $\times$ $128$. }
\label{fig:6}
\end{figure}
We find that the steady state snapshots of particle density Fig.\ref{fig:6} looks very much similar to the original velocity distribution we started with. Hence we say that the collection of active Brownian particles can mimic the substrate information. 
\section{discussion}
In this work we studied the  collection of active Brownian particles using coarse-grained and microscopic simulation on a two dimensional inhomogeneous substrate. The inhomogeneity is introduced as different pattern of activity particle experiences on the substrate. We looked for  different activity profiles: 
(i) Step function; (ii) Sigmoid; (iii) Gaussian; (iv) cone and finally we also studied the system with various asymmetric profiles: we studied the system with shapes 
of different alphabets. 
 We observed the patterns of local density, polarisation and current along the activity profile.  We find that for all four  distributions, the whole interface is symmetrically divided in four different quadrants.  Steady state is defined by such pattern of current. The conclusion is same for all four shapes. Interestingly density distributions are the mirror image of the activity profile. Starting from homogeneous density, in the steady state density achieves the lowest value in the high active region and then follow the pattern of the activity with sign reversed. We also performed the microscopic study and found that steady state density profile matches from the same as obtained by coarse-grained model. To check the same we replicate many {miscellaneous} shapes of activity for different alphabets of English, Hindi and Telugu. We find for all the cases in the steady state density follows the pattern of the activity. \\
Hence our results show the response of active Brownian particles on the substrate with inhomogeneous activity. Our results can be checked by the similar experiments as suggested in recent study of \cite{pellicciotta2023colloidal}. The results can be used to detect the pattern of the substrate as well as have  applications in printing using biological species or synthetic active particles.

\section{Author contributions}
The problem was designed by S.M. and numerically investigated by P.K.M. and A.K. All the authors analyzed and interpreted the results. The manuscript was prepared by contribution of each author. All the authors approved the final version of the manuscript.

 \section{acknowledgments}
PKM, AK and SM thanks PARAM Shivay for computatational facility under the National Supercomputing Mission, Government of India at the Indian Institute of Technology, Varanasi. Computing facility at Indian Institute of Technology(BHU), Varanasi is gratefully acknowledged.
S.M. thanks DST-SERB India, MTR/2021/000438, and
CRG/2021/006945 for financial support. PKM thanks University Grant Commision, India for providing scholarship.

 \nocite{*}
 \section{References}
\bibliography{references}

\end{document}